\documentclass[showpacs,aps,prl,twocolumn,groupedaddress]{revtex4}
\usepackage[dvips]{graphicx,color} %%% eliminate this line

\usepackage{graphicx}% Include figure files
\usepackage{dcolumn}% Align table columns on decimal point
\usepackage{bm}% bold math

\newcommand{\ee}{{$e^{+}e^{-}$}\ }
\newcommand{\kk}{{$K^{+}K^{-}$}\ }
\newcommand{\mev}{MeV/$c^2$}
\newcommand{\gev}{GeV/$c^2$}
\newcommand{\phiee}{$\phi \rightarrow e^{+}e^{-}$}
\newcommand{\rhoee}{$\rho \rightarrow e^{+}e^{-}$}
\newcommand{\omegaee}{$\omega \rightarrow e^{+}e^{-}$}
\newcommand{\omegaeepizero}{$\omega \rightarrow e^{+}e^{-}\pi^0$}
\newcommand{\etaee}{$\eta \rightarrow e^{+}e^{-}\gamma$}
\newcommand{\lambdappi}{$\Lambda \rightarrow$ p$\pi^-$}
\newcommand{\kspipi}{K$^{0}_s \rightarrow$ $\pi^+\pi^-$}
\newcommand{\shiftparm}{0.092}
\newcommand{\shiftparmerr}{0.002}
\newcommand{\shiftparmpct}{9}

\begin{document}

%\preprint{}

\title{Experimental Signature of Medium Modifications for $\rho$ and
$\omega$ Mesons\\ in the 12~GeV p~+~A Reactions}


\author{M.~Naruki}
 \email{naruki@nh.scphys.kyoto-u.ac.jp}

 \author{Y.~Fukao}
 \author{H.~Funahashi}
 \author{M.~Ishino}
 \altaffiliation[Present Address: ]{ICEPP, University of Tokyo, 7-3-1
 Hongo, Tokyo 113-0033, Japan} % second affiliation
 \author{H.~Kanda}
 \altaffiliation[Present Address: ]{Physics Department, Graduate School
 of Science, Tohoku University, Sendai 980-8578, Japan}
 \author{M.~Kitaguchi}
 \altaffiliation[Present Address: ]{Division of Quantum Beam Material
 Science,
 Kyoto University Research Reactor Institute,
 Kumatori-cho, Sennan-gun, Osaka 590-0494, Japan
 }
 \author{S.~Mihara}
 \altaffiliation[Present Address: ]{ICEPP, University of Tokyo, 7-3-1
 Hongo, Tokyo 113-0033, Japan}
 \author{K.~Miwa}
 \author{T.~Miyashita}
 \author{T.~Murakami}
 \author{T.~Nakura}
 \author{F.~Sakuma}
 \author{M.~Togawa}
 \author{S.~Yamada}
 \altaffiliation[Present Address: ]{Japan Atomic Energy Research
 Institute, Tokai, Ibaraki 319-1195, Japan.}
% E-mail address:yamada@neutrons.tokai.jaeri.go.jp}
 \author{Y.~Yoshimura}
 %\homepage{http://www-nh.scphys.kyoto-u.ac.jp/phi/}
\affiliation{%
Department of Physics, Kyoto University, Kitashirakawa Sakyo-ku, Kyoto
606-8502, Japan}%

 \author{H.~En'yo}
 \author{R.~Muto}
 \author{T.~Tabaru}
 \author{S.~Yokkaichi}
 \affiliation{RIKEN, 2-1 Hirosawa, Wako, Saitama 351-0198, Japan}%

\author{J.~Chiba}
 \altaffiliation[Present Address: ]{Faculty of Science and Technology, Tokyo
 University of Science, 2641 Yamazaki, Noda, Chiba 278-8510, Japan}%
\author{M.~Ieiri}
%\author{M.~Nomachi}
% \altaffiliation[Present Address: ]{Department of Physics, Osaka University, 1-1 Machikaneyama, Toyonaka, Osaka 560-0043, Japan.}%
\author{O.~Sasaki}
\author{M.~Sekimoto}
\author{K.~H. Tanaka}
\affiliation{
Institute of Particle and Nuclear Studies, KEK, 1-1 Oho, Tsukuba, Ibaraki 305-0801, Japan}%

 \author{H.~Hamagaki}
 \author{K.~Ozawa}
 \affiliation{Center for Nuclear Study, Graduate School of Science,
 University of Tokyo, 7-3-1 Hongo, Tokyo 113-0033, Japan}

%\thanks{}


\date{\today}

\begin{abstract}
 The invariant mass spectra of \ee pairs produced in 12-GeV
 proton-induced nuclear reactions
 are measured at the KEK Proton Synchrotron. On the low-mass side of the
 $\omega$ meson peak, a significant enhancement over the known hadronic
 sources has
 been observed. The mass spectra, including the excess, are well reproduced
 by a model that takes into account the density dependence of the vector
 meson mass modification, as theoretically predicted. 
\end{abstract}

%\keywords{}

\maketitle


%\section{\label{sec:level1}Introduction}
 It is well established that most of hadron masses are generated due to the
 spontaneous breaking of the chiral symmetry, which is the crucial aspect of
 the strong interaction.
 %As possible signatures for such a broken symmetry, 
 The modification of
 hadron masses and decay widths in hot and/or dense matter is
 theoretically predicted as a consequence of the restoration of the
 broken symmetry.
 Experimental observations of such phenomena have become one of the most
 interesting topics in hadron physics today.

 As for theoretical studies, 
%  the pioneering work of this field was done by Hatsuda and Kunihiro 20
%  years ago \cite{HK85}.
%  Since then,
 many works have been performed and they are summarized in \cite{HK94,
 brownrho}.
%, and recently in \cite{brownrho02}.
 Specifically, our experiment was motivated by two related studies
 concerning dense matter.
 Using an effective chiral Lagrangian,
 Brown and Rho proposed an in-medium scaling law, 
 which predicted a decrease of the vector meson mass at 20\% at
 the normal nuclear density $\rho_{0}$ \cite{br, brownrho02}.
 Hatsuda and Lee calculated the density dependence of the
 mass of vector mesons
 based on the QCD sum rule to reach the conclusion
 that the mass is approximately linear to the density in $0<\rho<2\rho_{0}$,
 and the mass of $\rho$ and $\omega$ mesons decreases at about 16\% at
 $\rho_{0}$ \cite{HL92, HLS95}.
% and in the range of $20$ to $40$ \mev\ for $\phi$ at $\rho_{0}$ \cite{HL92}.

 The experiment E325 has been performed at the KEK 12-GeV Proton Synchrotron
 to measure the invariant mass spectra of $\rho,\omega,\phi \rightarrow$
 \ee and $\phi \rightarrow $\kk decay modes simultaneously.
 Our main goal is to detect the modification of the spectral shape of
 vector mesons in nuclear media.
 In our earlier publication, we showed the significant enhancement
 below the $\omega$ peak in the invariant mass spectrum of \ee pairs
 for a copper target,  which is attributed to the mass-modification
 effect in nuclear matter \cite{pub0}.
 This was the first dilepton measurement of the modification of the vector
 meson mass in nucleus.

 It should be noted that our earlier and present results are
 strongly related to
 other few experimental observations. The CERES/NA45
 collaboration observed the low-mass electron-pair enhancement in Pb-Au
 collisions at 158 A GeV \cite{ceres}.
 The STAR collaboration reported the shift of $\rho$ meson peak in the
 $\pi^{+}\pi^{-}$ channel at RHIC in Au+Au and p+p collisions at
 $\sqrt{s}=$ 200 \gev\ \cite{star}.
 Their results may connote the chiral symmetry restoration in hot matter.
 The TAGX collaboration reported an in-medium modification of the $\rho$
 invariant mass distribution
 via $^{3}$He,$^{12}$C$(\gamma,\pi^{+}\pi^{-})$X reactions \cite{tagx}.
 Recently, the CBELSA/TAPS collaboration has reported that the mass
 modification of $\omega$ meson in the $\pi^{0}\gamma$ channel \cite{taps}.
%  Recently the GSI S236 Collaboration reported an
%  enhancement of the pion-nucleus potential in a deeply bound pionic
%  $^{115,119,123}$Sn nuclei as a possible signature of chiral symmetry
%  restoration \cite{pionic}. 

 Although these diverse experimental signatures, including ours,
 are not ample,
 they could be related to a critical behavior of hadrons in hot and/or
 dense matter. Compared to other experiments, our experiment is unique,
 since we are able to measure the
 \ee mass shapes directly with a high mass resolution and high
 statistics, placing a special emphasis on vector mesons that decay inside
 a nucleus. Needless to say,
 lepton pairs are almost free from final state interactions with nuclear
 matter.

 The results presented in this paper are based on most of the data that
 we have acquired. The statistics have improved by about 50-times
 compared to those given in a previous paper \cite{pub0}.
 We will show a comparison of the data to the theoretical
 prediction using a toy model dealing with the density dependence of the
 vector meson masses.

%\section{\label{sec:level2}Experimental Apparatus}
 A detailed description of the E325 spectrometer can be found elsewhere
 \cite{nim}. The detector elements, which are relevant for this
 analysis, are briefly discussed below.
 The spectrometer was designed to detect decays of slowly moving vector
 mesons, which have a larger probability to decay inside a target nucleus.
%  The spectrometer had two electron arms and two kaon arms,
%  which shared a dipole magnet and tracking devices.
 In 2002,
 the primary proton beam with a typical intensity of 7$\times$10$^8$ Hz was
 delivered to one carbon and four copper disk targets,
 which were aligned in-line along the beam axis.
 The interaction length of each copper target is
 0.054\%(73 mg/cm$^{2}$, 0.57\% in radiation length),
 and that of the carbon target is
 0.21\%(184 mg/cm$^{2}$, 0.43\% in radiation length).
%  One of the notable features of this experiment is that the high-intensity
%  primary beam was injected to the thin targets to reduce the background
%  due to $\gamma$ conversions in the target materials.
 The tracking system, consists of three drift chambers,
 gives the momentum resolution of
 $\sigma_{p}/p=\sqrt{(1.37\%\cdot p)^{2}+0.41\%^{2}}$ \cite{nim}.
%  The electron identification is performed with two or
%  three stages of iso-butane-filled gas-\v{C}erenkov counters and
%  lead-glass EM calorimeters in the momentum range of 0.4 to 2.7 GeV/$c$.
%  The lower limit is determined by the trigger condition in which
%  a geometrical matching of two stages of the electron ID counters was
%  required.
%  The upper limit is the momentum where pions started to
%  produce \v{C}erenkov light in iso-butane.
%  The electron detection efficiencies are typically 70\% with the
%  coincidence of two stages of gas-\v{C}erenkov counters and 87\% with
%  that of gas-\v{C}erenkov counters and EM calorimeters.

 The mass resolution and mass scale were examined for the observed peaks
 of \lambdappi\ and \kspipi\ decays.
 The observed resolutions and centroids of these resonances are consistent
 with the expectations given by a detailed detector simulation using Geant4
 \cite{g4}.
 The effects caused by
 multiple scattering, energy losses, the chamber resolution,
 and a miss-alignment of the tracking devices were minutely inspected.
 Please note that the mass spectra presented in this manuscript is not
 corrected for the mass scale which is 3.7$\pm$0.8 \mev\ lower than the
 real value at 0.8\gev\ (i.e. $\omega$ peak).
 At the low-mass side of the resonances, a long tail arises due to the
 energy loss of electrons, mainly caused by bremsstrahlung.
 Such a tail is estimated to be
 negligible compared to the excess that we have been observing.
 The mass resolutions for \omegaee\ and \phiee\ decays are estimated to
 be 8.0 and 10.7 \mev, respectively.
 These values are consistent with the data when we neglected the excess
 parts of the mass spectra.

%\section{\label{sec:level3}Results}
\begin{figure}[htbp]
 \includegraphics[width=8.0cm]{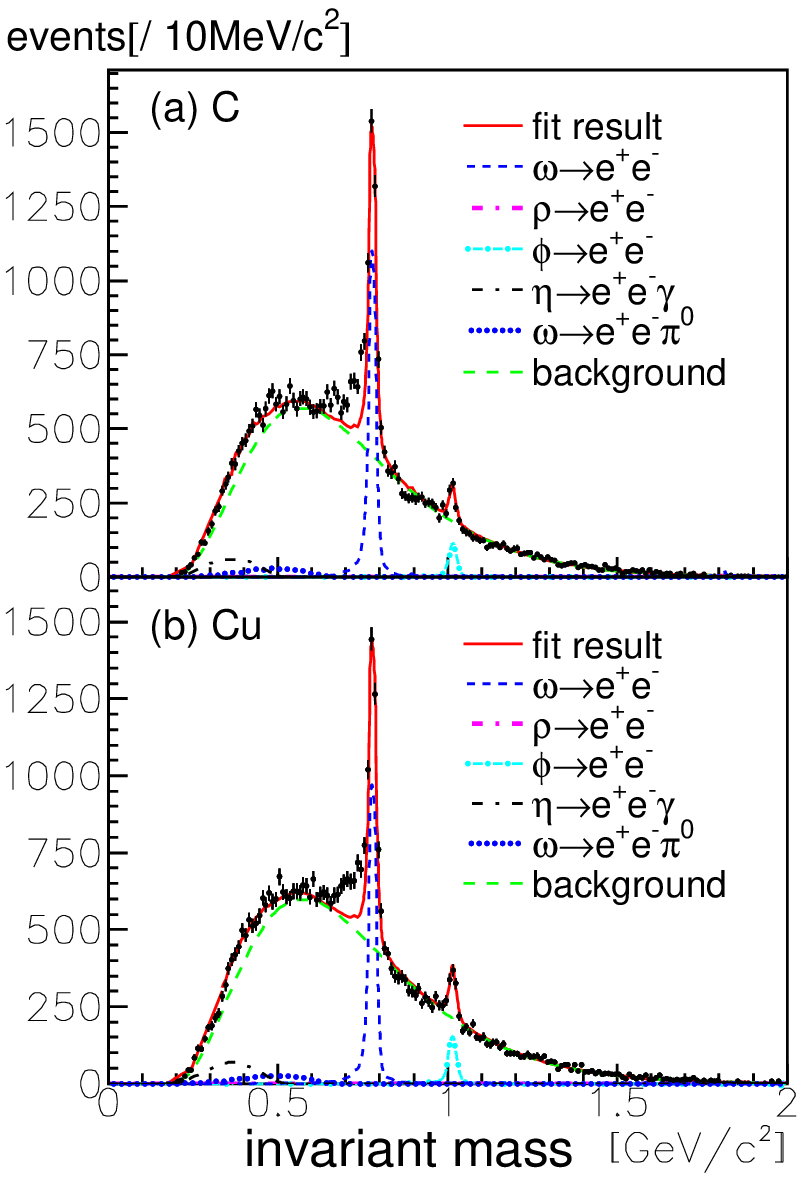}\\
 \caption{\label{fig:ee}
 Invariant mass spectra of $e^{+}e^{-}$ for the (a) C and
 (b) Cu targets. The solid lines
 are the best-fit results, which is the sum of the known hadronic
 decays, \omegaee\ (dashed), \phiee\ (thick dash-dotted),
 \etaee\ (dash-dotted), and \omegaeepizero\ (dotted)
 together with the combinatorial background (long-dashed).
 \rhoee\ is not visible (see text).}
\end{figure}

 Figure \ref{fig:ee} shows the invariant mass spectra
 of the \ee  pairs using all of the data taken in 2002.
 We have required each track of an \ee pair to
 come into each of the two arms; therefore,
 the low-mass region of the spectra is largely suppressed.
 The invariant mass spectra was fitted with the combinatorial
 background and known hadronic sources: \rhoee,
 \omegaee, \phiee, \etaee, and \omegaeepizero.
 The combinatorial background was evaluated by the event-mixing method.
%  The contribution to the background due to an electron miss identification
%  was estimated to be less than \chk{13} \%.
% Spectral shapes of hadrons were obtained with the nuclear cascade code,
% JAM \cite{jam}, to take the mass acceptance into account.
 The relativistic Breit-Wigner distribution was used to obtain
 the spectral shapes of resonances.
 The mass resolution and the detector effect of our spectrometer were
 taken into account through the detector simulation described before.
 The kinematical distributions of mesons have been obtained by the nuclear
 cascade code JAM \cite{jam}, which is in a good agreement with the real
 data.
 The relative abundances of these components were determined by the
 fitting, except for the ratio of \omegaeepizero\ to \omegaee\ decay which
 was fixed to their branching ratios, 59/6.95 given by the PDG \cite{pdg}.
 The fit results are plotted with the solid lines in Fig.\ref{fig:ee}
 and summarized in Table \ref{tab:yield}.
 The obtained $\chi^{2}/d.o.f.$ are 161/140 and 154/140 for the C and Cu
 targets, respectively.
 The region from 0.6 to 0.76 \gev\ was excluded from the fit,
 because the fit including this region resulted in failure
 at C.L. 99.9\%
 as listed in Table \ref{tab:yield2} 
 \footnote{We have also examined the Gounaris-Sakurai shape \cite{gs} and the
 relativistic Breit-Wigner shape multiplied by the Boltzmann factor
 \cite{boltz}. These shapes do not reproduce the data at C.L.99.9\%.}.
 \begin{table}
  \caption{Signal yields in the acceptance
  together with $\chi^{2}/d.o.f.$ for the C and Cu targets,
  obtained by the fit excluding the mass range of 0.6 to 0.76\gev.
  The values for $\rho$ are expressed as an upper limit at 95\% C.L.
  }\label{tab:yield}
 \begin{ruledtabular}
 \begin{tabular}{@{}c@{ }c@{ }c@{ }c@{ }c@{ }c@{ }c@{}}
  & $\eta$ Dalitz & $\phi$ & $\omega$ & $\rho$ & excess & $\chi^{2}$/dof\\\hline
 C&1012$\pm$112&398$\pm$42&3644$\pm$92&(112)&1461$\pm$131&161/140\\
Cu&1249$\pm$126&547$\pm$45&3346$\pm$91&(169)&1341$\pm$136&154/140\\
  \end{tabular}
  \caption{\label{tab:yield2}Same as Table \ref{tab:yield} but
  corresponds to the fit including the excess region.}
 \begin{tabular}{@{}c@{ }c@{ }c@{ }c@{ }c@{ }c@{ }c@{}}
  & $\eta$ Dalitz & $\phi$ & $\omega$ & $\rho$ & excess & $\chi^{2}$/dof\\\hline
 C&1025$\pm$116&347$\pm$40&3552$\pm$98&1468$\pm$242&893$\pm$169&366/162\\
Cu&1248$\pm$132&505$\pm$44&3209$\pm$98&1405$\pm$257&798$\pm$175&295/162\\
  \end{tabular}
 \end{ruledtabular}
 \end{table}

\begin{figure*}[bht]
 \includegraphics[width=11cm]{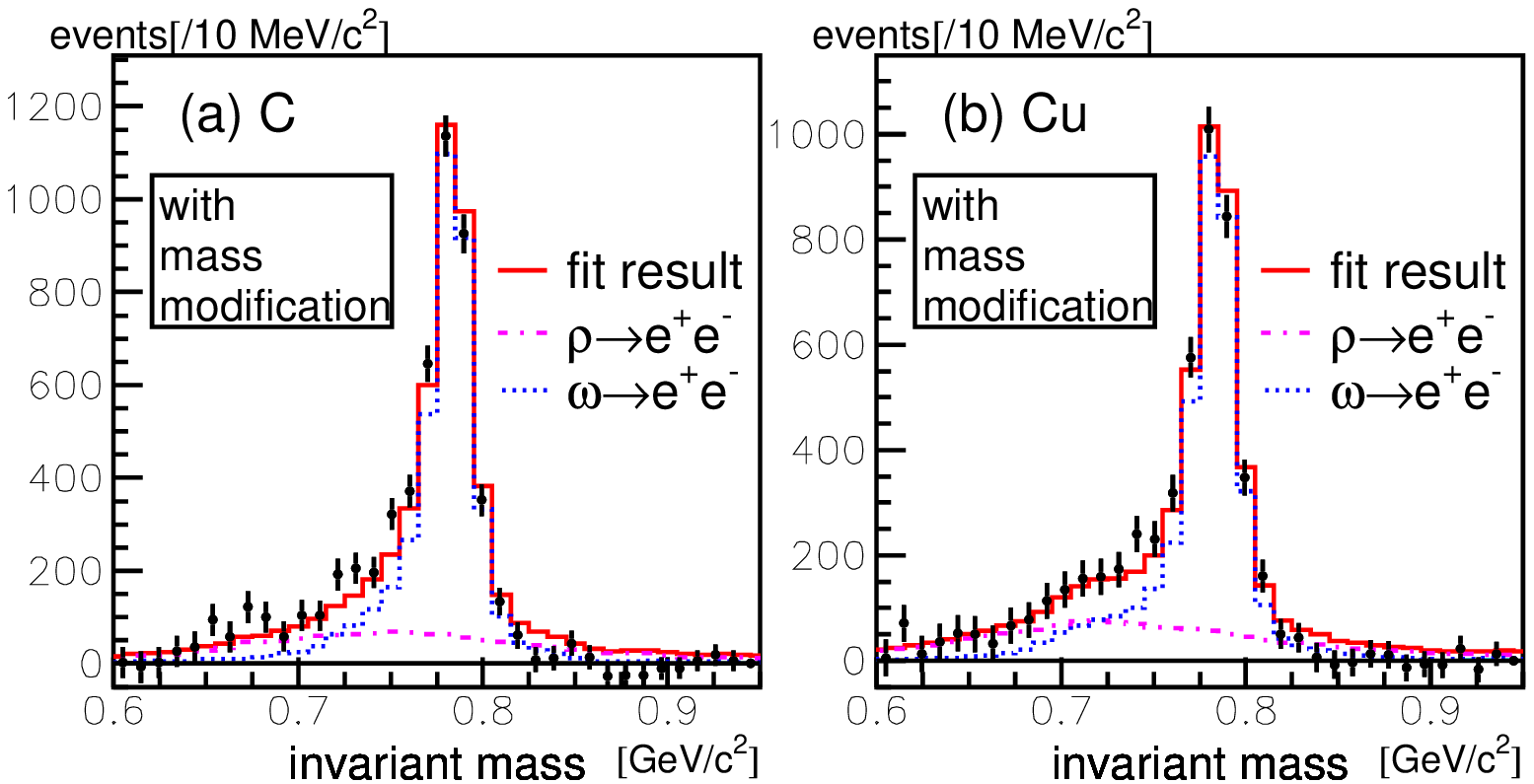}\hspace*{1cm}%
 \includegraphics[width=5.5cm]{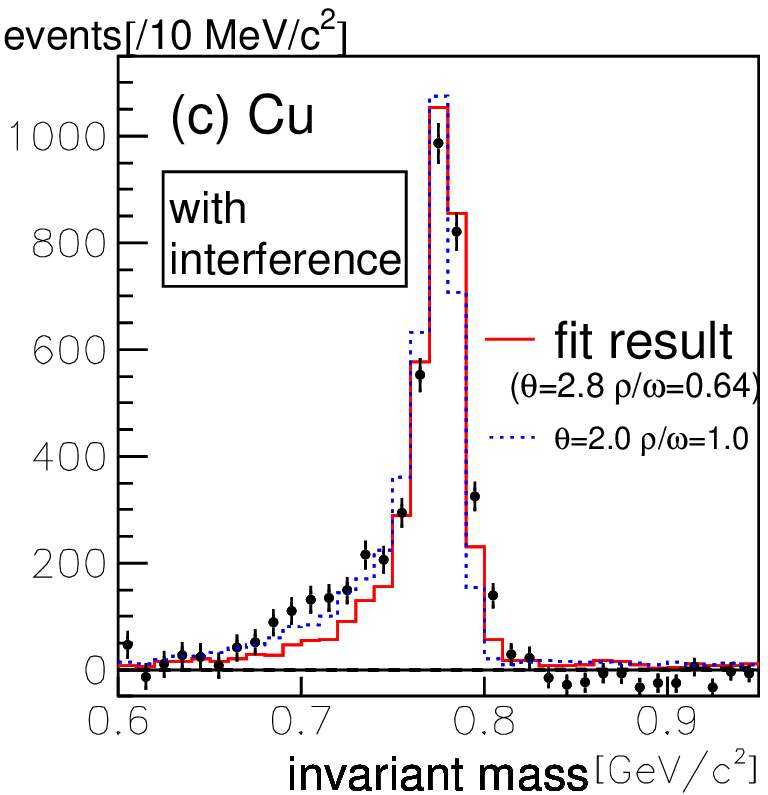}
 \caption{\label{fig:mc}
 Invariant mass spectra of $e^{+}e^{-}$.
 The combinatorial background and the shapes of \etaee\ and
 \omegaeepizero\ were subtracted.
 The result of the model calculation considering the in-media
 modification for the (a) C and (b) Cu targets,
 together with
 (c) the fit result with the $\rho-\omega$ interference for the Cu target.
 The solid lines show the best-fit results.
 In (a) and (b),
 the shapes of \omegaee\ (dotted) and \rhoee\ (dash-dotted)
 were modified according to the model using the
 formula $m(\rho)/m(0) = 1 - k(\rho/\rho_{0})$ with $k$ = \shiftparm.
 In (c), the solid line shows the $\rho-\omega$ interfering shape with the
 $\rho/\omega$ = 0.64 and the interference angle = 2.8 radian.
 The dotted line is the fit result with the typical values of the
 $\rho/\omega$ ratio = 1.0 and the angle = 2.0 radian \cite{rob}.
}
\end{figure*}

 A significant excess can be seen on the low-mass side of the
 $\omega$ peak, whereas the high-mass tail of the $\omega$ can be
 reproduced with the expected shapes.
 The number of excess was evaluated by subtracting the amplitude for the
 fit function from the data, in the range of 0.6 to 0.76\gev.
 The obtained ratios of the excess to the $\omega$ peak
 in the acceptance
 are estimated to be 0.40$\pm$0.04 and 0.40$\pm$0.04 for the C and Cu
 targets, respectively.
 The obtained yields of each resonance and the excess are given in Table
 \ref{tab:yield}.
%  was estimated to be 1408$\pm$123(1273$\pm$124) for the carbon(copper)
%  target. 
 %The ratio of $\rho$ to $\omega$ can be determined by the fit.
 This procedure determines the ratio of $\rho$ to $\omega$,
 provided that they follow the relativistic Breit-Wigner distribution
 without any mass modification.
 After the acceptance correction,
 the 95\% C.L. allowed parameter regions are obtained as
 $\rho/\omega<$ 0.15 and % 0.03(stat.) * 1.96 sigma + 0.09(sys.)
 $\rho/\omega<$ 0.31 % 0.05(stat.) * 1.96 sigma + 0.21(sys.)
 for C and Cu targets, respectively.
 The systematic errors arises from the uncertainty of the background
 estimation, which amount to 0.09 and 0.21 for the C and Cu targets, respectively.
 The obtained $\rho/\omega$ ratios are much smaller than unity, as was
 previously measured in $pp$ interactions at the same energy \cite{blobel}.
 A possible explanation is that
 most of the $\rho$ are decaying inside the nuclei due to their short
 lifetime; their mass is modified in nuclear media and contribute to
 the excess.

 A comparison has been done for the data with a model considering the
 in-medium mass modification.
 In this model, the mass decreases linearly as a function of the
 density $\rho$, in the following relation:
  $m(\rho)/m(0) \simeq 1 - k(\rho/\rho_{0})$ \cite{HL92,HLS95}.
 The parameter $k$ was expected to be 16$\pm$6\% for $\rho$ and $\omega$ meson,
 where $\rho_{0}$ is the normal nuclear density \cite{HLS95}.
 The pole mass was modified by the above formula according
 to the density at the decay point.
 The decay-width modification was neglected.
 A Woods-Saxon shape was used for the nuclear density distribution;
 $\rho/\rho_{0}\propto(1+exp((r-R)/\tau))^{-1}$,
 where $R=2.3$ and $4.1$ fm, $\tau=0.57$ and $0.50$ fm for the C and Cu
 targets, respectively.
 We assumed that the vector mesons were generated at the surface of an
 incident hemisphere of the target nucleus.
 This assumption is reasonable, since we have observed 
 the mass-number dependence of the $\omega$ production cross section
 as $\sigma(A)\propto A^{2/3}$ \cite{tabaru}.
 This model predicts that the probabilities of $\rho$ meson decays inside
 a nucleus are 46\% and 61\% for the C and Cu targets, respectively,
 while those of $\omega$ are 5\% and 9\%.

\begin{figure}[bht]
 \includegraphics[width=8.0cm,keepaspectratio]{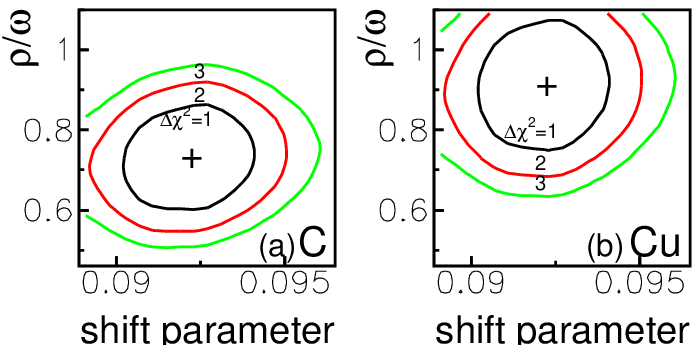}
 \caption {\label{fig:chi}
 Confidence ellipsoids
 for the shift parameter $k$ (see text),
 and the production ratio $\rho/\omega$,
 for the (a) C and (b) Cu targets.
 The step of each ellipsoid corresponds to $\Delta\chi^{2}$=1.
 The best fit point is indicated by the cross.
 }
\end{figure}

 Based on the model, we modified the shapes of $\rho$ and $\omega$,
 by introducing the shift parameter $k$ which is common for the C and Cu
 targets. We fit again the entire mass region of Fig.\ref{fig:ee} using
 the same procedure as before.
 The fit results,
 after subtracting the combinatorial background and the shapes of
 \etaee\ and \omegaeepizero\,
 are shown in Fig.\ref{fig:mc}(a) and (b).
 The spectra for both C and Cu targets can be reproduced quite well
 by this model.
 Confidence ellipsoids for the shift parameter $k$, and the
 $\rho/\omega$ ratio are also shown in Fig.\ref{fig:chi}.

 We obtained the shift parameter of $k$ = \shiftparm$\pm$\shiftparmerr.
 The best-fit values of the $\rho/\omega$ ratio are 0.7$\pm$0.1 and
 0.9$\pm$0.2 for the C and Cu targets, respectively.
 Please note that here the yield of $\rho$ and $\omega$ includes both
 free decays and in-medium decays.
 It is concluded that the
 observed modification can be understood with the model in which the masses of
 the $\rho$ and $\omega$ mesons decrease by \shiftparmpct\% at the
 normal nuclear density.
 This value is consistent with the theoretical prediction \cite{HLS95}.

 We have neglected in-media mass broadening, which is theoretically 
 predicted in \cite{asakawa, herrmann, KKW, peters} and well summarized
 in \cite{RW, post},
 since it appears that such width broadening of $\rho$ and $\omega$,
 which increases the yield of the high-mass tail, does not fit our data
 \footnote{Here we assumed that the width broadening is proportional to the
 density, the fit results actually favored the zero-broadening case.}.

 It is suggested that the $\rho-\omega$ interference is
 a possible explanation of the observed spectral modification \cite{rob}.
 Such possibility has been inspected by fitting the data with the
 interfering shape,
 which depends on the production ratio $\rho/\omega$,
 and the interference angle $\theta$ \cite{horn}.
 We scanned the $\rho/\omega$ ratio from 0.2 to 2.6,
 and $\theta$ from 0.0 to 3.6 radian.
 It turned out that the fit was unsuccessful for any combination
 in the scanned regions, and rejected at $99.9\%$ C.L.
 Around the suggested interference angle, 2 radian \cite{rob},
 the minimum $\chi^{2}$ 
 was found when $\theta$ was 2.8 and the $\rho/\omega$ ratio was 0.64.
 This result is shown in Fig.\ref{fig:mc}(c), together with the curve
 for $\theta$ = 2.0 radian and the $\rho/\omega$ ratio = 1.0 as a reference.

 In summary,
 we have observed the excess over the known hadronic sources on the
 low-mass side of the $\omega$ meson peak in the \ee invariant mass spectra.
 This demonstrates the spectral-shape modification of $\rho$ and
 $\omega$ mesons in nuclear matter.
 Without the mass modification of hadrons,
 the obtained $\rho$/$\omega$ ratio, which is consistent with zero,
 contradicts to the known $\rho/\omega$ ratio at this energy.
 A possible explanation is
 that most of the $\rho$ mesons decay inside the
 nucleus, and contribute to the excess.
 The observed excess is understood by the model in which the mass of
 $\rho/\omega$ meson decreases by \shiftparmpct\% at the normal nuclear
 density, whereas the $\rho-\omega$ interference can not explain the data.
\begin{acknowledgments}
 We would like to thank Prof. T. Hatsuda for helpful discussions.
 We greatly acknowledge all the staff members of KEK-PS,
 especially the beam channel group for their helpful support.
 This work was partly funded by 
 the Japan Society for the Promotion of Science,
 RIKEN Special Postdoctoral Researchers Program and
 a Grant-in-Aid for Scientific Research from the Japan Ministry
 of Education, Culture, Sports, Science and Technology (MEXT).
 Finally, we also thank the staff members of RIKEN Super Combined Cluster
 system and RIKEN-CCJ.
\end{acknowledgments}

\end{document}